\documentclass[preprint,aps,pra,showpacs,floatfix,superscriptaddress]{revtex4-1}
\usepackage{longtable}
\usepackage{graphicx}
\usepackage{feynmp}
\usepackage{amsmath,amsfonts,amssymb,verbatim,float}
\usepackage[usenames]{color}    %for colours
\usepackage{colortbl}
\usepackage{calrsfs}
\usepackage{ulem} % for remarks
\usepackage{hyperref} % for links in the bibliography
\usepackage{subfigure}
\usepackage{indentfirst}
\usepackage{tikz} %for plots
\usepackage{pgfplots}
\graphicspath{{graph/}}
\definecolor{faded}{gray}{0.45}
\newcommand{\bp}{{\bf p}}
\newcommand{\bk}{{\bf k}}
\newcommand{\br}{{\bf r}}
\newcommand{\bb}{{\bf b}}

\newcommand{\balpha}{\boldsymbol{\alpha}}

\newcommand{\bepsilon}{\boldsymbol{\epsilon}}
\begin{document}
\thispagestyle{empty}
\title{
Bremsstrahlung from twisted electrons in the field of heavy nuclei
}
\author{M.~E.~Groshev}
\affiliation{
Department of Physics, St. Petersburg State University,
Universitetskaya Naberezhnaya 7/9, 199034 St. Petersburg, Russia
}
\author{V.~A.~Zaytsev}
\affiliation{
Department of Physics, St. Petersburg State University,
Universitetskaya Naberezhnaya 7/9, 199034 St. Petersburg, Russia
}
\author{V.~A.~Yerokhin}
\affiliation{
Center for Advanced Studies, Peter the Great St.~Petersburg Polytechnic University,
St. Petersburg 195251, Russia
}
\author{V.~M.~Shabaev}
\affiliation{
Department of Physics, St. Petersburg State University,
Universitetskaya Naberezhnaya 7/9, 199034 St. Petersburg, Russia
}
%
% BEGIN ABSTRACT ==============================================================
%
\begin{abstract}
We present a fully relativistic calculation of the bremsstrahlung emitted by twisted electrons propagating in the field of bare heavy nuclei. 
The electron-nucleus interaction is accounted for to all orders in the nuclear binding strength parameter $\alpha Z$, 
thus allowing us to investigate the bremsstrahlung in a strong field, 
where the effects of the ``twistedness'' are expected to be most pronounced.
To explore these effects, we study the angular and polarization properties of the photons emitted in course of the inelastic twisted electrons scattering by the gold target.
The influence of the kinematic parameters of the incident electrons on the double-differential cross section and the degree of the linear polarization is also discussed.
\end{abstract}
%
% END ABSTRACT ================================================================
%
\maketitle
%
%
%-------------------------------------------------------------------------------
%
\section{INTRODUCTION}
%
%-------------------------------------------------------------------------------
%
%BEGIN INTRODUCTION===========================================================
%
The so-called twisted (or vortex) electrons presently attract considerable interest from both experimental and theoretical sides (see~\cite{Bliokh_PR690_1:2017, Lloyd_RMP89_035004_2017, Larocque_CP59_126_2018} for a review and relevant references). 
The interest is partly caused by the fact that the twisted particles can carry a large
%, when compared to the conventional (plane-wave) electrons,
total angular momentum (TAM) projection $\hbar m$ onto the propagation direction.
% Nowadays, the twisted electrons with $m \sim 1000$ can be readily produced in experiments~\cite{Mafakheri}.
In particular, twisted electrons with $m \sim 1000$ can be readily produced with current experimental techniques~\cite{Mafakheri}.
The magnetic dipole moment of such electrons $m\mu_B$ ($\mu_B$ is the Bohr magneton) is by three orders of magnitude larger than one of the plane-wave electrons.
As a result, the role of the magnetic interaction in processes involving vortex electron beams is significantly enhanced.
This makes twisted electrons a very promising tool for studying magnetic properties of different materials and surfaces~\cite{Rusz_PRL111_105504:2013, Beche_NatPhys10_26:2014, Schattschneider_Ultra136_81:2014, Edstrom_PRL116_127203:2016} and for detecting various subtle magnetic effects~\cite{Ivanov_2013}.
% The electrons that possess a well-defined total angular momentum (TAM) projection onto the propagation direction are called twisted. Shortly after the theoretical prediction~\cite{Bliokh_PRL99_190404:2007} of these electron states the experimental realizations of twisted electrons have followed~\cite{Verbeeck_N497_301:2010, Uchida_N464_737:2010, McMorran_S331_192:2011}. Since then, the investigations dedicated to the study of the interactions of these particles with various surfaces and materials are of rapidly growing interest (see~\cite{Bliokh_PR690_1:2017, Lloyd_RMP89_035004_2017, Larocque_CP59_126_2018} for a review and relevant references). The interest in these studies comes from ***
%
% - - - - - - - - - - - - - - - - - - - - - - - - - - - - - - - - - - - - - - - 
%
\\ \indent
%
% - - - - - - - - - - - - - - - - - - - - - - - - - - - - - - - - - - - - - - - 
%
All these and other possible applications of twisted electrons are based on properties of fundamental processes of interaction between the twisted beams and atomic targets.
Theoretical descriptions of basic atomic processes involving vortex electrons are, therefore, highly demanded.
Up to now such descriptions were presented for the radiative recombination~\cite{Matula_NJP16_053024_2014, Zaytsev_PRA95_012702},  
elastic scattering~\cite{Boxem_PRA89_032715_2014, Serbo_PRA92_012705_2015, Karlovets_PRA92_052703_2015, Karlovets_PRA95_032703_2017, Kosheleva},  impact excitation~\cite{Boxem_PRA91_032703_2015}, and impact ionization~\cite{Harris_JPB52_094001_2019}.
The theoretical investigation of the bremsstrahlung from twisted electrons, however, has not yet been reported so far.
In the present paper, we aim to fill this gap by developing the fully relativistic description of this process.
We follow the formalism developed in Ref.~\cite{Zaytsev_PRA95_012702} and describe an incoming twisted electron as a coherent superposition of the conventional (plane-wave) electrons propagating in a central potential.
As a result, the amplitude of the process is expressed as a coherent sum of the amplitudes of the bremsstrahlung from the plane-wave electrons whose evaluation can be performed with the usage of the well-developed techniques~\cite{TsengPratt, Jakubassa_PRA82_042714:2010, Yerokhin_PRA82_062702:2010, Muller_PRA90_032707:2014, Mangiarotti_RPC141_312:2017, Jakubassa_PRA100_032703:2019}.
% In this approach, the interaction of the incoming and outgoing electrons with the central potential of the target is accounted for nonperturbatively. That allows one to describe the bremsstrahlung from twisted electrons in the field of the heavy systems.
This approach, accounting for the interaction of the incoming and outgoing electrons with the central potential of the target nonperturbatively, allows one to describe the bremsstrahlung from twisted electrons in the field of heavy atoms.
Heavy systems are of particular importance for studies of effects of ``twistedness'' since one may expect a strong enhancement of these effects due to large spin-orbit interaction. 
%The interaction of the vortex electrons with such systems is of particular interest. 
%Indeed, the influence of the ``twistedness'' of the incoming electron is expected to be the most pronounced in heavy systems, where the spin-orbit interaction is strongly enhanced.
% In these systems, where the role of the spin-orbit interaction is strongly enhanced, the effects of the ``twistedness'' are expected to be the most pronounced.
%
% - - - - - - - - - - - - - - - - - - - - - - - - - - - - - - - - - - - - - - - 
%
\\ \indent
%
% - - - - - - - - - - - - - - - - - - - - - - - - - - - - - - - - - - - - - - - 
%
The paper is organized as follows: 
In Sec.~\ref{brem_from_pl} we recall basic relations for the bremsstrahlung from conventional (plane-wave) electrons. 
The theoretical description of the bremsstrahlung from twisted electrons is presented in Sec.~\ref{brem_from_tw}. 
In Secs.~\ref{DDCS} and \ref{Stokes_parameters} the numerical results for the double-differential cross section and the Stokes parameters are presented, respectively. 
Finally, a summary and an outlook are given in Sec.~\ref{conclusion}.
%
% - - - - - - - - - - - - - - - - - - - - - - - - - - - - - - - - - - - - - - - 
%
\\ \indent
%
% - - - - - - - - - - - - - - - - - - - - - - - - - - - - - - - - - - - - - - - 
%
Relativistic units ($m_e = \hbar = c = 1$) and the Heaviside charge units ($e^2 = 4\pi\alpha$) are utilized in the present paper.
%
% END INTRODUCTION =============================================================
%-------------------------------------------------------------------------------
%
\section{BASIC FORMALISM}
%
%-------------------------------------------------------------------------------
%
The description of the bremsstrahlung from twisted electrons can be given in terms of the formulas derived for the plain-wave electrons. 
We, therefore, start with the compilation of the basic properties of the bremsstrahlung from conventional (plane-wave) electrons.
We apply the approach based on the relativistic partial-wave decomposition of the electron's wave function and the multipole expansion of the photon's wave function.
This approach is the most appropriate for the description of the bremsstrahlung in the field of heavy systems~\cite{TsengPratt, Yerokhin_PRA82_062702:2010} where the interaction between the electron and nucleus should be accounted for nonperturbatively.
% (see, e.g.~\cite{TsengPratt, Jakubassa_PRA82_042714:2010, Yerokhin_PRA82_062702:2010, Muller_PRA90_032707:2014, Mangiarotti_RPC141_312:2017, Jakubassa_PRA100_032703:2019}). 
%
%
%
%-------------------------------------------------------------------------------
%
\subsection{Bremsstrahlung from conventional (plane-wave) electrons}
\label{brem_from_pl}
%
%-------------------------------------------------------------------------------
%
The probability of an emission of a photon with the four-momentum $(\omega,\bk)$ 
and the polarization $\lambda$ in the process of the inelastic scattering of an electron with the four-momentum $(\varepsilon_i,\bp_i)$ 
and the helicity $\mu_i$ from the bare nucleus is given by~\cite{Akhiezer_1965, Landau_IV}
\begin{equation}
\label{eq:plane_probability}
dW^{\rm (pl)}_{\mu_f,\lambda;\bp_i\mu_i} 
= 
(2\pi)^4 
|\tau^{\rm (pl)}_{\mu_f,\lambda;\bp_i\mu_i}|^2 
\delta(\varepsilon_i-\varepsilon_f -\omega)
d\bk d\bp_f.
\end{equation}
Here $\varepsilon_f$, $\bp_f$ and $\mu_f$ are the energy, the asymptotic momentum, and the helicity of the outgoing electron, respectively. 
The amplitude of the bremsstrahlung expresses as 
\begin{equation}
\tau^{\rm (pl)}_{\mu_f,\lambda;\bp_i\mu_i}  = 
\int d\br 
\Psi^{(-)\dagger}_{\bp_f \mu_f}(\br) 
\hat R_{\bk \lambda}^\dagger(\br) 
\Psi^{(+)}_{\bp_i \mu_i}(\br),
\label{eq:amplitude}
\end{equation}
with the photon emission operator given by
\begin{equation}
\hat{R}^{\dagger}_{\bk\lambda}(\br)
 = 
-\sqrt{\frac{\alpha}{(2\pi)^2\omega}}
\balpha\cdot{\bepsilon}_{\lambda}^*
e^{-i\bk\cdot \br},
\label{eq:photon_emission}
\end{equation}
where $\balpha$ stands for the vector of Dirac matrices and the Coulomb gauge defines the polarization vector.
The wave functions of the incoming $\Psi_{\bp_i\mu_i}^{(+)}$ and outgoing $\Psi_{\bp_f\mu_f}^{(-)}$ electrons express as follows~\cite{Rose_1961, Pratt_RMP45_273:1973, Eichler_1995}
\begin{equation}
\Psi_{\bp\mu}^{(\pm)}(\br)
 = 
\frac{1}{\sqrt{4\pi p\varepsilon}} 
\sum\limits_{\kappa m_j}
C^{j\mu}_{l0\,1/2\mu}
i^l 
\sqrt{2l+1}
e^{\pm i \delta_{\kappa}}
D^j_{m_j\mu}(\varphi_{\hat{\bp}}, \theta_{\hat{\bp}}, 0)
\Psi_{\varepsilon \kappa m_j}(\br).
\label{eq:wf_in_out}
\end{equation}
Here $p = |\bp|$, 
$\kappa = (-1)^{l+j+1/2}\left(j + 1/2\right)$ is the Dirac quantum number 
determined by the angular momentum $j$ and the parity $l$, 
$C^{JM}_{j_1 m_1\,j_2 m_2}$ is the Clebsch-Gordan coefficient, 
$\delta_{\kappa}$ is the phase shift induced by the scattering central potential, 
$D^J_{MM'}$ is the Wigner matrix~\cite{Rose_1957, Varshalovich}, 
the azimuthal $\varphi_{\hat{\bp}}$ and polar $\theta_{\hat{\bp}}$ angles define the direction of the unit vector $\hat{\bp}$, 
and $\Psi_{\varepsilon \kappa m_j}$ is the partial-wave Dirac solution. 
%
% - - - - - - - - - - - - - - - - - - - - - - - - - - - - - - - - - - - - - - - 
%
\\ \indent
%
% - - - - - - - - - - - - - - - - - - - - - - - - - - - - - - - - - - - - - - - 
%
Substituting Eq.~\eqref{eq:wf_in_out} into Eq.~\eqref{eq:amplitude} and utilizing the well-known multipole expansion for the photon emission operator~\eqref{eq:photon_emission}, we obtain the final expression in the form of the triple expansion over the electron partial waves and photon multipoles.
%the triple-infinite sum over the electron partial waves and photon multipoles.
%
The partial amplitudes, i.e. terms of this sum, are evaluated by separating the angular integration and performing it analytically.
The remaining integral over the radial variable has to be calculated numerically (see, e.g., Ref.~\cite{Yerokhin_PRA82_062702:2010}).
Summing over the partial-waves and multipoles up to desired accuracy, we determine the bremsstrahlung amplitude which defines the probability~\eqref{eq:plane_probability}, and, consequently, all the properties of the process studied.
%
%-------------------------------------------------------------------------------
%
\subsection{Bremsstrahlung from twisted electrons}
\label{brem_from_tw}
\begin{figure}[h]
\includegraphics{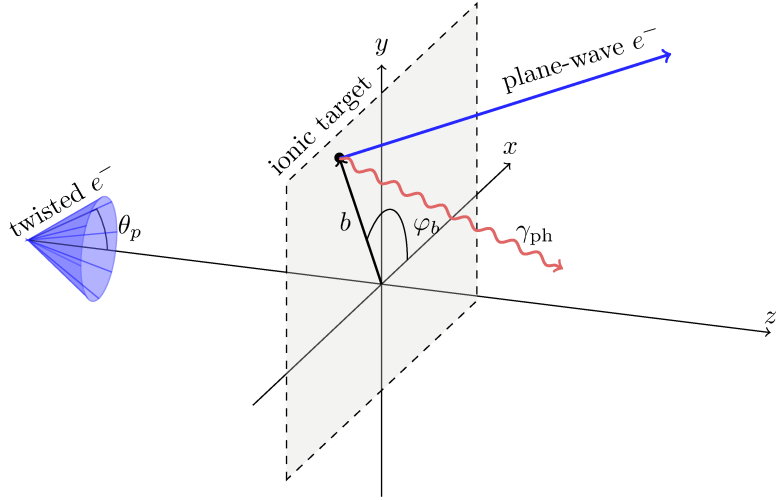}
\caption{
The geometry of the bremsstrahlung from twisted electrons in the field of a single bare nucleus shifted from the $z$-axis by the impact parameter $\bb$.
}
\label{brem_geom_single_target}
\end{figure}

\begin{figure}[h]
\includegraphics{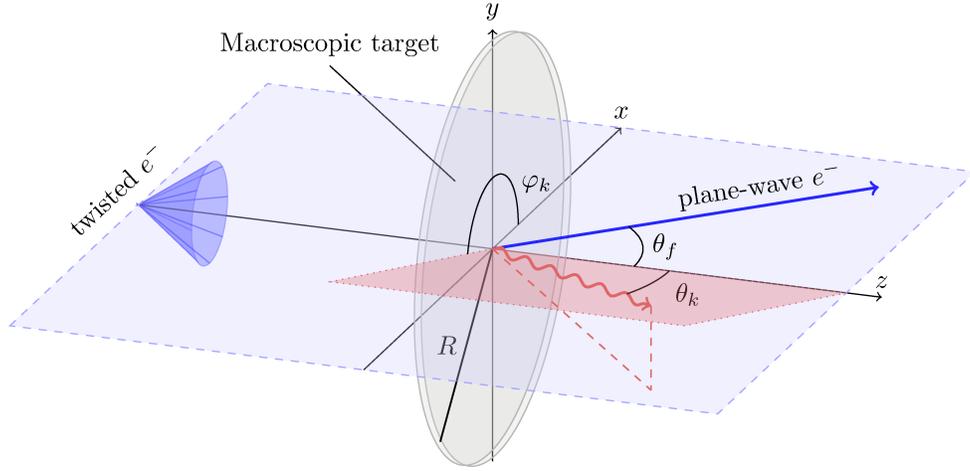}
\caption{
Geometry of the bremsstrahlung process from twisted electrons scattered by a macroscopic target.
}
\label{fig:geom_macr}
\end{figure}

%
%-------------------------------------------------------------------------------
%
The theoretical description of the free twisted electrons is well represented in the literature 
(see~\cite{Bliokh_PR690_1:2017, Lloyd_RMP89_035004_2017, Larocque_CP59_126_2018} 
for a review and relevant references).
Here we only briefly sketch the important properties of these electrons, 
which are taken in the form of a Bessel wave in the present work.
Twisted electrons possess the well-defined energy $\varepsilon$, 
helicity $\mu$, 
and the total angular $m$ and linear $p_z$ momenta projections onto the same direction.
The $z$-axis is fixed along this direction.
In the momentum space, 
these states represent a cone with the opening angle $\theta_p = \arctan\left(\varkappa/p_z\right)$ 
where $\varkappa = \sqrt{\varepsilon^2 - 1 - p_z^2}$ stands for the well-defined transversal momentum.
The inhomogeneity of the probability distribution 
and the flux density with respect to the space variable 
is one of the distinguishing features of the twisted electrons.
This feature results in the dependence of the scattering process on the relative position of the target and the vortex beam.
%
% The twisted electrons, in contrast to the plane-wave ones, have the inhomogeneous probability distribution and the flux density.
% As a result, the relative position of the target and the vortex beam is important.
%
% - - - - - - - - - - - - - - - - - - - - - - - - - - - - - - - - - - - - - - - 
%
\\ \indent
%
% - - - - - - - - - - - - - - - - - - - - - - - - - - - - - - - - - - - - - - - 
%
We start with the consideration of the bremsstrahlung from twisted electrons in the field of a single bare nucleus, which is shifted from the $z$-axis by the impact parameter $\bb = (b_x, b_y, 0)$ as shown in Fig.~\ref{brem_geom_single_target}.
%
%
%%%%%%%%%%%%%%%%%%%%
%%%%%%%%%%%%%%%%%%%%
%
Here and below we assume that both the emitted photon 
and the outgoing electron are asymptotically described by the plane waves.
This corresponds to the assumption that detectors used in the actual experiments do not register ``twistedness" of the particles, which is the case for the present-day experimental setups. 
%This assumption reflects the fact that nowadays the detectors of the conventional (plane-wave) particles are mostly utilized.
%This assumption corresponds to the fact that in present day experiments the detectors of the conventional (plane-wave) particles are mostly utilized.
%
The probability of the process depicted in Fig.~\ref{brem_geom_single_target} is given by
\begin{equation}
dW^{\rm (tw)}_{\mu_f, \lambda; \varkappa m p_z \mu_i}(\bb)
 = 
(2\pi)^4 n_i 
|\tau^{\rm (tw)}_{\mu_f, \lambda; \varkappa m p_z \mu_i}(\bb)|^2 
\delta(\varepsilon_i-\varepsilon_f -\omega)
d{\bf k} d{\bf p}_f,
\label{eq:tw_prob}
\end{equation}
with the scattering amplitude expressing as
\begin{eqnarray}
\nonumber
\tau^{\rm (tw)}_{\mu_f, \lambda;\varkappa m p_z\mu_i}(\bb)
& = &
\int d\br 
\Psi^{(-)}_{\bp_f \mu_f}(\br-\bb)
\hat{R}^{\dagger}_{\bk \lambda}(\br-\bb)
\Psi^{(+)}_{\varkappa m p_{z}\mu_i}(\br)
\\ & = &
\int d\br
\Psi^{(-)}_{\bp_f \mu_f}(\br)
\hat{R}^{\dagger}_{\bk \lambda}(\br)
\Psi^{(+)}_{\varkappa m p_{z}\mu_i}(\br+\bb).
\label{eq:amplitude_tw_0}
\end{eqnarray}
The explicit form of the wave function of the asymptotically twisted electron propagating in the central field can be found in Ref.~\cite{Zaytsev_PRA95_012702}:
\begin{equation}
\Psi^{(+)}_{\varkappa m p_{z}\mu_i}(\br+\bb) 
 =
 \int d\bp
\frac{e^{im\varphi_p}}{2\pi p_{\perp}}
\delta(p_{\perp}-\varkappa)\delta(p_{\parallel}-p_z)
i^{\mu_i-m}
\Psi^{(+)}_{{\bp}\mu_i}({\br})
e^{i\bp\cdot\bb},
\label{eq:wf_tw}
\end{equation}
where $p_\perp$ and $p_\parallel$ are the transverse and longitudinal components of the momentum $\bp$, respectively.
Substituting Eq.~\eqref{eq:wf_tw} into Eq.~\eqref{eq:amplitude_tw_0} one can express the amplitude of the bremsstrahlung from twisted electron through the one appearing in the plane-wave case~\eqref{eq:amplitude}
\begin{equation}
\tau^{\rm (tw)}_{\mu_f, \lambda;\varkappa m p_z \mu_i}({\bb}) 
= 
\int d\bp
\frac{e^{im\varphi_p}}{2\pi p_{\perp}}
\delta(p_{\perp}-\varkappa)\delta(p_{\parallel}-p_z)
i^{\mu_i-m}
e^{i\bp\cdot\bb}
\tau_{\bp_f\mu_f,\bk\lambda;\bp\mu_i}^{\rm (pl)}.
\label{eq:amplitude_tw}
\end{equation}
The scattering amplitude~\eqref{eq:amplitude_tw} defines uniquely the probability of the process under consideration, from which all measurable quantities can be determined.
%
%Therefore, the theoretical description of the bremsstrahlung from twisted electrons is regarded as complete.
%
% - - - - - - - - - - - - - - - - - - - - - - - - - - - - - - - - - - - - - - - 
%
\\ \indent
%
% - - - - - - - - - - - - - - - - - - - - - - - - - - - - - - - - - - - - - - - 
%
Up to now we have considered the bremsstrahlung from twisted electrons in a field of a single ionic or atomic target.
The experimental investigation of such a process is a challenging task.
We now consider a more realistic scenario, namely, the bremsstrahlung from the twisted electrons scattered by the infinitely extended (macroscopic) target.
We describe this target as an incoherent superposition of ions (or atoms) being randomly and homogeneously distributed.
The fully differential cross section of this scenario being schematically depicted in Fig.~\ref{fig:geom_macr} is given by~\cite{Serbo_PRA92_012705:2015}
\begin{equation}
\frac{d\sigma^{\rm (tw)}_{\mu_f,\lambda;\varkappa p_z\mu_i}}{d\omega d\Omega_k d\Omega_f}
= 
\int \frac{d\bb}{\pi R^2}
\left\vert
\tau^{\rm (tw)}_{\mu_f, \lambda;\varkappa m p_z\mu_i}(\bb)
\right\vert^2
=
\frac{1}{\cos\theta_p} \int \frac{d\varphi_p}{2\pi}
\frac{d\sigma^{\rm (pl)}_{\mu_f, \lambda;\bp\mu_i}}{d\omega d\Omega_k d\Omega_f},
\label{eq:twisted_tdcs}
\end{equation}
where $\left(\pi R^2\right)$ is the cross section area with $R$ being the radius of the cylindrical box and 
\begin{equation}
\frac{d\sigma^{\rm (pl)}_{\mu_f, \lambda;\bp_i\mu_i}}{d\omega d\Omega_k d\Omega_f}
=
\frac{1}{v_i}
\frac{dW^{\rm (pl)}_{\mu_f, \lambda; \bp_i\mu_i}}{d\omega d\Omega_kd\Omega_f}
=
(2\pi)^4 \omega^2 \frac{p_f \varepsilon_f}{v_i}
|\tau^{\rm (pl)}_{\mu_f, \lambda; \bp_i\mu_i}|^2
\label{eq:plane_tdcs}
\end{equation}
is the fully differential cross section of the bremsstrahlung from plane-wave electrons with the velocity $v_i$.
From Eq.~\eqref{eq:twisted_tdcs} it is seen that in the case of the scattering of the twisted electron by the macroscopic target the fully differential cross section does not depend on $m$.
%
% - - - - - - - - - - - - - - - - - - - - - - - - - - - - - - - - - - - - - - - 
%
\\ \indent
%
% - - - - - - - - - - - - - - - - - - - - - - - - - - - - - - - - - - - - - - - 
%
In the present investigation, 
we restrict ourselves to the consideration of the scenario 
in which the incoming twisted electron is unpolarized and only the emitted photon is detected.
This process is described by the double-differential cross section (DDCS)
\begin{equation}
\label{eq:ddcs}
d\sigma_\lambda \equiv
\frac{d\sigma^{\rm (tw)}_{\lambda;\varkappa p_z}}{d\omega d\Omega_k}
=
\frac{1}{2}\sum_{\mu_i\mu_f}
\int d\Omega_f 
\frac{d\sigma^{\rm(tw)}_{\mu_f, \lambda; \varkappa p_z\mu_i}}{d\omega d\Omega_k d\Omega_f}
=
\frac{1}{\cos\theta_p}
\int \frac{d\varphi_p}{2\pi}
\frac{d\sigma^{\rm (pl)}_{\lambda;\bp}}{d\omega d\Omega_k},
\end{equation}
where
\begin{equation}
\frac{d\sigma^{\rm (pl)}_{\lambda;\bp_i}}{d\omega d\Omega_k}
=
\frac{1}{2}\sum_{\mu_i\mu_f}
\int d\Omega_f 
\frac{d\sigma^{\rm (pl)}_{\mu_f, \lambda; \bp_i\mu_i}}{d\omega d\Omega_k d\Omega_f}
\end{equation}
is the DDCS for the bremsstrahlung from plane-wave electrons.
%
%
%===============================================================================
%
\section{RESULTS AND DISCUSSIONS}
%
%===============================================================================
%
As already has been mentioned the effects of the ``twistedness'' are expected to be most pronounced for heavy systems.
We consider, therefore, the bremsstrahlung from twisted electrons scattered by the gold target.
Having in mind that the bremsstrahlung from electrons with $\varepsilon_i \gtrsim 100$ keV comes mostly from the nuclei, we here consider the scattering of twisted electrons by an infinite macroscopic target consisting of bare gold nuclei.
%In particular, the scattering by the pure Coulomb potential of the bare gold nuclei is studied.
This choice of our model allows us to explicitly demonstrate the effects induced by the ``twistedness'' of the incident electron.
% Such restriction is expected to be applicable for the incoming electron energies chosen in the present investigation, viz. $100$ and $500$ keV.
% 
%-------------------------------------------------------------------------------
%
\subsection{Double differential cross section}
\label{DDCS}
\begin{figure}[h!]
\includegraphics[width=\textwidth]{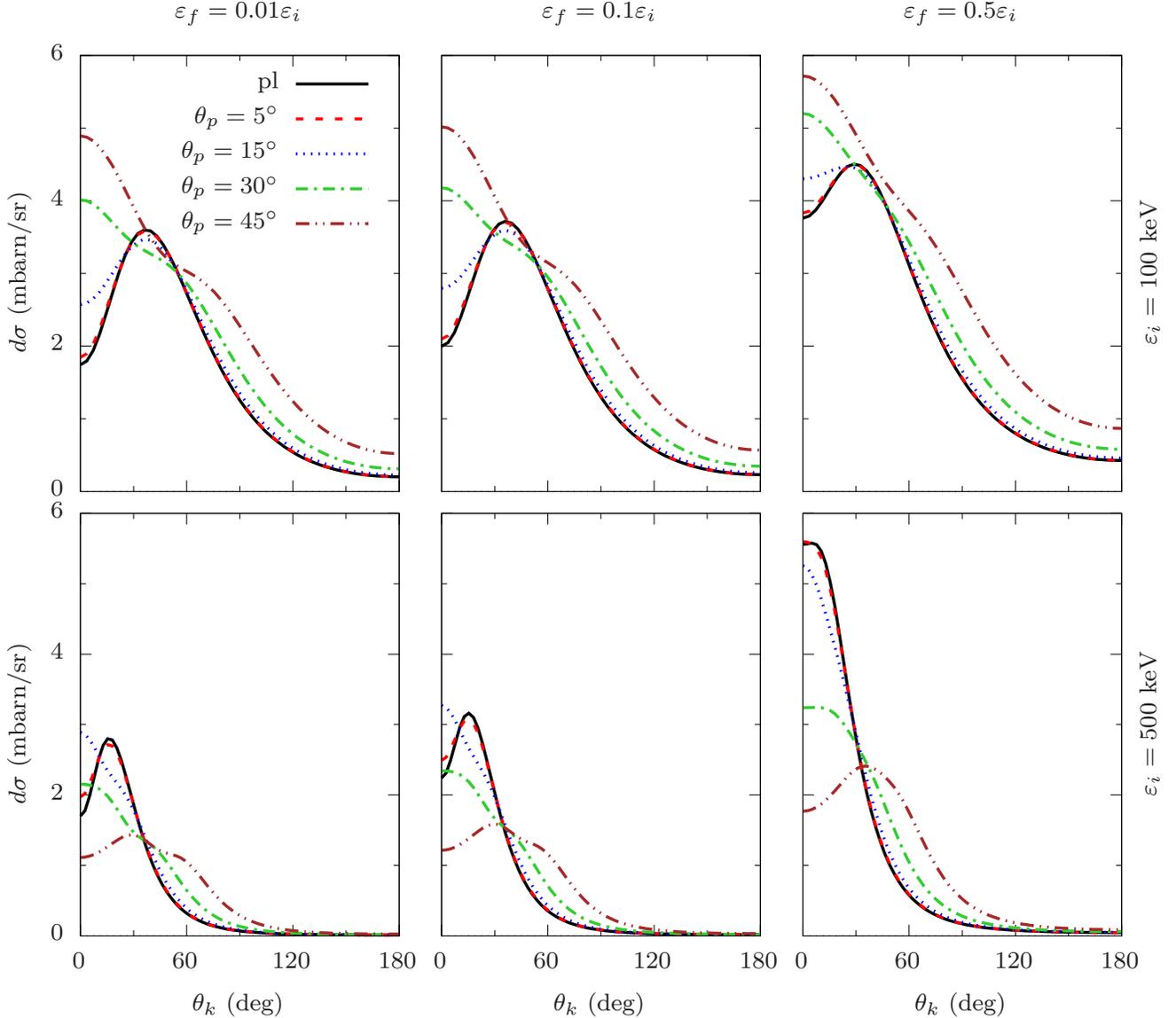}
\caption{
The scaled DDCS $d\sigma \equiv (\omega/Z^2)\sum_\lambda d\sigma_\lambda$ for
the bremsstrahlung from the twisted electrons with the opening (conical) angle $\theta_p$ and the energy $\varepsilon_i = 100$ keV (first row) and $500$ keV (second row) as a function of the photon emission angle $\theta_k$.
The left, middle, and right panels correspond to the energies of the outgoing electron $\varepsilon_f = 0.01\varepsilon_i$, $0.1\varepsilon_i$, and $0.5\varepsilon_i$, respectively. 
}
\label{fig:ddcs}
\end{figure}

%
%-------------------------------------------------------------------------------
%
We start from the analysis of the DDCS for the bremsstrahlung from $100$ keV and $500$ keV twisted electrons scattered by the macroscopic target consisting of bare gold ($Z = 79$) nuclei.
Figure~\ref{fig:ddcs} presents the DDCS~\eqref{eq:ddcs} summed over $\lambda$ for three different energies of the outgoing electron $\varepsilon_f = 0.01\varepsilon_i$, $0.1\varepsilon_i$, and $0.5\varepsilon_i$.
From this figure, it is seen that the growth of the opening angle $\theta_p$ leads to the strong qualitative changes in the angular distribution of the bremsstrahlung. 
These changes, however, manifest differently for different energies of the incident electron.
Indeed, for $\varepsilon_i = 100$ keV (first row in Fig.~\ref{fig:ddcs}) the probability of the forward bremsstrahlung increases with the growth of the opening angle and at $\theta_p \gtrsim 30^\circ$ the DDCS turns into a monotonically decreasing function with a maximum at $\theta_k = 0^{\circ}$. 
In contrast to this, for $\varepsilon_i = 500$ keV (second row in Fig.~\ref{fig:ddcs}) the increase of the opening angle leads to the drop of the probability of the forward bremsstrahlung.
The most significant change of the DDCS for this incident electron energy happens at $\varepsilon_f = 0.5$ and $\theta_p = 45^\circ$.
% Let us note that for this incident electron energy the most significant change of the DDCS is expected for $\varepsilon_f = 0.5$ and $\theta_p = 45^\circ$
For these parameters, the formation of the a maximum at the photon emission angles $\theta_k$ from $30^\circ$ to $60^\circ$ is predicted.
%
%------------------------------------------------------------------------------
\subsection{Stokes parameters}
\label{Stokes_parameters}
\begin{figure}[h!]
\includegraphics[width=\textwidth]{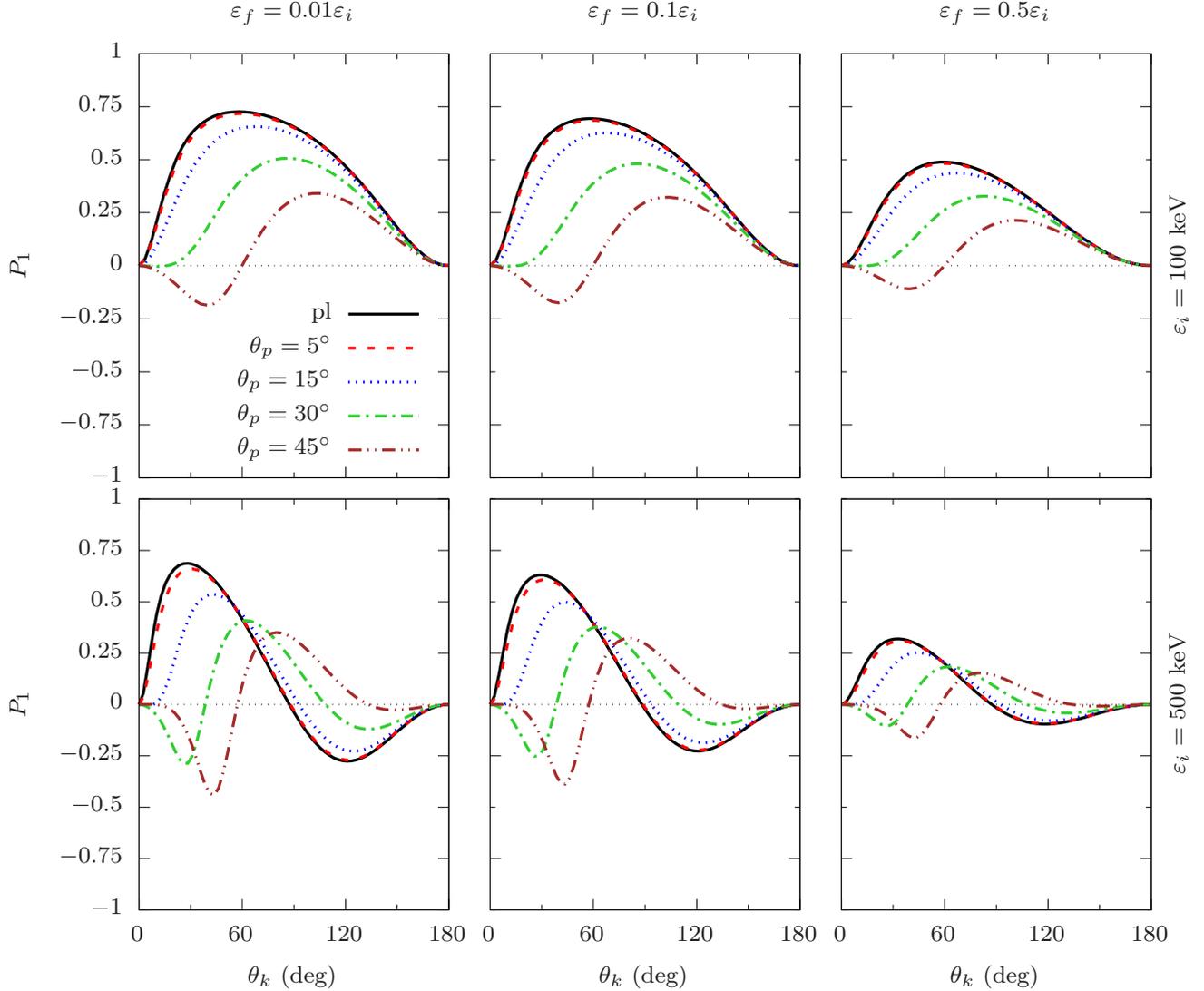}
\caption{
The degree of the linear polarization (the first Stokes parameter $P_1$) for the bremsstrahlung from the twisted electrons with the opening (conical) angle $\theta_p$ and the energy $\varepsilon_i = 100$ keV (first row) and $500$ keV (second row) as a function of the photon emission angle $\theta_k$.
The left, middle, and right panels correspond to the energies of the outgoing electron $\varepsilon_f = 0.01\varepsilon_i$, $0.1\varepsilon_i$, and $0.5\varepsilon_i$, respectively.
}
\label{fig:stokes}
\end{figure}

%------------------------------------------------------------------------------
%
We now turn to the investigation of the ``twistedness''-induced effects on the polarization properties of the bremsstrahlung.
For this purpose, we evaluate the Stokes parameters which are defined as
% We now turn to the analysis of the polarization properties of the bremsstrahlung from twisted electrons. This analysis can be naturally performed by evaluating the Stokes parameters which are given by
%
\begin{equation}
\label{eq:Stokes}
P_1 = \frac{d\sigma_{0^{\circ}} - d\sigma_{90^{\circ}}}{d\sigma_{0^{\circ}} + d\sigma_{90^{\circ}}},
\quad
P_2 = \frac{d\sigma_{45^{\circ}} - d\sigma_{135^{\circ}}}{d\sigma_{45^{\circ}} + d\sigma_{135^{\circ}}},
\quad
P_3 = \frac{d\sigma_{+1} - d\sigma_{-1}}{d\sigma_{+1} + d\sigma_{-1}}.
\end{equation}
Here 
$d\sigma_\lambda$ is the DDCS expressed by Eq.~\eqref{eq:ddcs} and 
\begin{equation}
d\sigma_{\chi}  
\equiv 
\frac{d\sigma^{\rm (tw)}_{\chi;\varkappa p_z}}{d\omega d\Omega_k}
\end{equation}
is the DDCS for the emission of the photon with the linear polarization 
 characterized by the angle $\chi$ 
\begin{equation}
\bepsilon_\chi 
= 
\frac{1}{\sqrt{2}}
\sum\limits_{\lambda=\pm1}
e^{-i\lambda\chi}\bepsilon_\lambda.
\end{equation}
%
% - - - - - - - - - - - - - - - - - - - - - - - - - - - - - - - - - - - - - - - 
%
\\ \indent
%
% - - - - - - - - - - - - - - - - - - - - - - - - - - - - - - - - - - - - - - - 
%
Figure~\ref{fig:stokes} presents the degree of the linear polarization (the first Stokes parameter $P_1$) 
for the bremsstrahlung from twisted electrons with energy $\varepsilon_i = 100$ keV and $500$ keV, 
scattered by a macroscopic target consisting of bare gold ($Z = 79$) nuclei. 
We note that for the process under investigation $P_2$ and $P_3$ are identically equal to zero. 
From Fig.~\ref{fig:stokes} it is seen that, like in the case of DDCS (see Fig.~\ref{fig:ddcs}), 
the degree of the linear polarization exhibits a strong dependence on the kinematic parameters of the incident twisted electron. 
As an example, for the $100$ keV incident electron energy (first row in Fig.~\ref{fig:stokes}) and $\theta_p = 45^\circ$, 
the first Stokes parameter $P_1$ takes negative values. 
This corresponds to the bremsstrahlung which is linearly polarized perpendicular to the scattering plane. 
A similar effect was predicted for the Vavilov-Cherenkov radiation by twisted electrons~\cite{Ivanov_PRA93_053825:2016} 
and for the radiative recombination of twisted electrons~\cite{Zaytsev_PRA95_012702}.
For $\varepsilon_i = 500$ keV (second row in Fig.~\ref{fig:stokes}) 
we observe a more pronounced dependence on the opening (conical) angle $\theta_p$.
Already at $\theta_p$ from $15^\circ$ to $30^\circ$, the degree of the linear polarization undergoes qualitative changes when compared to the plane-wave case.
Such a strong dependence of the first Stokes parameter $P_1$ on the opening angle allows one 
to consider the bremsstrahlung process as a possible tool for the diagnostics of the twisted electron beams.
%
%-------------------------------------------------------------------------------
%
\section{CONCLUSION}
\label{conclusion}
%
%-------------------------------------------------------------------------------
%
In this work we have developed a fully-relativistic description of the bremsstrahlung by twisted electrons in the field of heavy nuclei. 
Our approach accounts for the interaction of the incoming vortex and outgoing plane-wave electrons with the Coulomb field of the target to all orders in the nuclear binding strength parameter $\alpha Z$.
%
% - - - - - - - - - - - - - - - - - - - - - - - - - - - - - - - - - - - - - - - 
%
\\ \indent
%
% - - - - - - - - - - - - - - - - - - - - - - - - - - - - - - - - - - - - - - - 
%
The developed approach has been applied for the investigation of the angular and polarization properties of the bremsstrahlung from twisted electrons scattered by the macroscopic target consisting of bare gold nuclei.
In particular, the double-differential cross section and the degree of linear polarization $P_1$ of the bremsstrahlung as functions of the photon emission angle were evaluated for different energies of the outgoing electron.
It was found that both the DDCS and $P_1$ do depend strongly on the kinematic properties of the twisted electron, namely, the opening angle $\theta_p$ and the energy $\varepsilon_i$.
Additionally, the dependence on $\theta_p$ is qualitatively different for different incident electron energies.
As an example, for $\varepsilon_i = 100$ keV the forward bremsstrahlung increases with the growth of $\theta_p$ , while for $\varepsilon_i = 500$ keV it decreases.
It was also found that $P_1$ exhibits a stronger dependence on $\theta_p$ for $500$ keV twisted electrons when compared to the $100$ keV ones.
%
%----------------------------------------------------------------------------------------------------------------------
%
\section{ACKNOWLEDGMENTS}
%
%----------------------------------------------------------------------------------------------------------------------
%
This work was supported by the grant of the President of the Russian Federation (Grant No. MK-4468.2018.2), by RFBR (Grant No. 18-32-00602), and by SPbSU-DFG (Grants No. 11.65.41.2017 and No. STO 346/5-1). 
V.~A.~Y. acknowledges the support by the Ministry of Education and Science of the Russian Federation Grant No. 3.5397.2017/6.7. V.~M.~S. acknowledges the support from the Foundation for the advancement of theoretical physics and mathematics ``BASIS''.
The work of V.~A.~Z. was also supported by SPbSU (TRAIN2019\_2: 41159446).
%
%----------------------------------------------------------------------------------------------------------------------
%BEGIN BIBLIOGRAPHY==============================================

%END BIBLIOGRAPHY=================================================

\begin{thebibliography}{99}
%
% \bibitem{Bliokh_PRL99_190404:2007}
% K.~Y.~Bliokh, Y.~P.~Bliokh, S.~Savel\'ev, and F.~Nori,
% {\it Phys. Rev. Lett.} {\bf 99}, 190404 (2007).
%
% \bibitem{Verbeeck_N497_301:2010}
% J.~Verbeeck, H.~Tian, and P.~Schattschneider,
% {\it Nature} {\bf 467}, 301 (2010).
%
% \bibitem{Uchida_N464_737:2010}
% M.~Uchida and A.~Tonomura,
% {\it Nature} {\bf 464}, 737 (2010).
%
% \bibitem{McMorran_S331_192:2011}
% B.~J.~McMorran, A.~Agrawal, I.~M.~Anderson, 
% A.~A.~Herzing, H.~J.~Lezec, J.~J.~McClelland, and J.~Unguris,
% {\it Science} {\bf 331}, 192 (2011).
%
\bibitem{Bliokh_PR690_1:2017}
K.~Y.~Bliokh, I.~P.~Ivanov, G.~Guzzinati, L.~Clark, R.~Van~Boxem, A.~B\'ech\'e, R.~Juchtmans, M.~A.~Alonso, P.~Schattschneider, F.~Nori, and J.~Verbeeck,
Phys. Rep. {\bf 690}, 1 (2017).
%
\bibitem{Lloyd_RMP89_035004_2017}
S.~M.~Lloyd, M.~Babiker, G.~Thirunavukkarasu, and J.~Yuan,
Rev. Mod. Phys. {\bf 89}, 035004 (2017).
%
\bibitem{Larocque_CP59_126_2018}
H.~Larocque, I.~Kaminer, V.~Grillo, G.~Leuchs, M.~J.~Padgett, R.~W.~Boyd, M.~Segev, and E.~Karimi,
Contemp. Phys. {\bf 59}, 126 (2018).
%
\bibitem{Mafakheri} 
E.~Mafakheri, A.~H.~Tavabi, P.-H.~Lu, R.~Balboni, F.~Venturi, C.~Menozzi,
G.~C.~Gazzadi, S.~Frabboni, A.~Sit, R.~E.~Dunin-Borkowski, E.~Karimi,
and V.~Grillo,
Appl. Phys. Lett. {\bf 110}, 093113 (2017).
%
\bibitem{Rusz_PRL111_105504:2013}
J.~Rusz, and S.~Bhowmick,
Phys. Rev. Lett. {\bf 111}, 105504 (2013).
%
\bibitem{Beche_NatPhys10_26:2014}
A.~Beche, R.~Van~Boxem, G.~V.~Tendeloo, and J.~Verbeeck,
Nat. Phys. {\bf 10}, 26 (2014).
%
\bibitem{Schattschneider_Ultra136_81:2014}
P.~Schattschneider, S.~Löffler, M.~Stöger-Pollach, and J.~Verbeeck,
Ultramicroscopy {\bf 136}, 81 (2014).
%
\bibitem{Edstrom_PRL116_127203:2016}
A.~Edström, A.~Lubk, and J.~Rusz,
Phys. Rev. Lett. {\bf 116}, 127203 (2016).
%
\bibitem{Ivanov_2013}
I.~P.~Ivanov and D.~V.~Karlovets, 
Phys. Rev. Lett. 110, 264801 (2013); Phys. Rev. A 88 043840 (2013).
%
\bibitem{Matula_NJP16_053024_2014}
O.~Matula, A.~G.~Hayrapetyan, V.~G.~Serbo, A.~Surzhykov, and S.~Fritzsche,
New J. Phys. {\bf 16}, 053024 (2014).
%
\bibitem{Zaytsev_PRA95_012702}
V.~A.~Zaytsev, V.~G.~Serbo, and V.~M.~Shabaev, 
Phys. Rev. A {\bf 95}, 012702 (2017).
%
\bibitem{Boxem_PRA89_032715_2014}
R.~V.~Boxem, B.~Partoens, and J.~Verbeeck,
Phys. Rev. A {\bf 89}, 032715 (2014).
%
\bibitem{Serbo_PRA92_012705_2015}
V.~G.~Serbo, I.~P.~Ivanov, S.~Fritzsche, D.~Seipt, and A.~Surzhykov,
Phys. Rev. A {\bf 92}, 012705 (2015).
%
\bibitem{Karlovets_PRA92_052703_2015}
D.~V.~Karlovets, G.~L.~Kotkin, and V.~G.~Serbo,
Phys. Rev. A {\bf 92}, 052703 (2015).
%
\bibitem{Kosheleva}
V.~P.~Kosheleva, V.~A.~Zaytsev, A.~Surzhykov, V.~M.~Shabaev, and Th.~St\"ohlker, 
Phys. Rev. A {\bf 98}, 022706 (2018). 
%
\bibitem{Karlovets_PRA95_032703_2017}
D.~V.~Karlovets, G.~L.~Kotkin, and V.~G.~Serbo, and A.~Surzhykov,
Phys. Rev. A {\bf 95}, 032703 (2017).
%
\bibitem{Boxem_PRA91_032703_2015}
R.~V.~Boxem, B.~Partoens, and J.~Verbeeck,
Phys. Rev. A {\bf 91}, 032703 (2015).
%
\bibitem{Harris_JPB52_094001_2019}
A.~L.~Harris, A.~Plumadore, and Z.~Smozhanyk,
J. Phys. B {\bf 52}, 094001 (2019).
%
\bibitem{TsengPratt}
H.~K.~Tseng and R.~H.~Pratt, 
Phys. Rev. A {\bf 3}, 100 (1971); 
Phys. Rev. A {\bf 7}, 1502 (1973); 
Phys. Rev. Lett. {\bf 33}, 516 (1974); 
 Phys. Rev. A {\bf 19}, 1525 (1977).
%
\bibitem{Jakubassa_PRA82_042714:2010}
D.~H.~Jakubassa-Amundsen, 
Phys. Rev. A {\bf 82}, 042714 (2010).
%
\bibitem{Yerokhin_PRA82_062702:2010}
V.~A.~Yerokhin and A.~Surzhykov, 
Phys. Rev. A {\bf 82}, 062702 (2010).
%
\bibitem{Muller_PRA90_032707:2014}
R.~A.~M\"uller, V.~A.~Yerokhin and A.~Surzhykov, 
Phys. Rev. A {\bf 90}, 032707 (2014).
%
\bibitem{Mangiarotti_RPC141_312:2017}
A.~Mangiarotti and M.~N.~Martins, 
Rad. Phys. Chem. {\bf 141}, 312 (2017).
%
\bibitem{Jakubassa_PRA100_032703:2019}
D.~H.~Jakubassa-Amundsen, 
Phys. Rev. A {\bf 100}, 032703 (2019).
%
\bibitem{Akhiezer_1965}
A.~I.~Akhiezer and V.~B.~Berestetskii, 
{\it Quantum Electrodynamics} (Interscience, New York, 1965).
%
\bibitem{Landau_IV}
V.~B.~Berestetsky, E.~M.~Lifshitz, and L.~P.~Pitaevskii, 
{\it Quantum Electrodynamics} (Butterworth-Heinemann, Oxford, 2006).
%
\bibitem{Rose_1961}
M.~E.~Rose, 
{\it Relativistic Electron Theory}, (Wiley, New York, 1961).
%
\bibitem{Pratt_RMP45_273:1973}
R.~H.~Pratt, A.~Ron, and H.~K.~Tseng, 
Rev. Mod. Phys. {\bf 45}, 273 (1973); {\bf 45}, 663(E) (1973).
%
\bibitem{Eichler_1995}
J.~Eichler and W.~Meyerhof, 
{\it Relativistic Atomic Collisions} (Academic, San Diego, 1995).
%
\bibitem{Rose_1957}
M.~E.~Rose,
{\it Elementary Theory of Angular Momentum}, (Wiley, New York, 1957).
%
\bibitem{Varshalovich}
D.~A.~Varshalovich, A.~N.~Moskalev, and V.~K.~Khersonskii,
{\it  Quantum Theory of Angular Momentum}, 
(World Scientific, Singapore, 1988).
%
\bibitem{Serbo_PRA92_012705:2015}
V.~G.~Serbo, I.~P.~Ivanov, S.~Fritzsche, D.~Seipt, and A.~Surzhykov, 
Phys. Rev. A {\bf 92}, 012705 (2015).
%
\bibitem{Ivanov_PRA93_053825:2016}
I.~P.~Ivanov, V.~G.~Serbo, and V.~A.~Zaytsev,
Phys. Rev. A {\bf 93}, 053825 (2016).
%
\end{thebibliography}
\end{document}